\documentclass[
    prd,
    reprint,
    amsmath,
    amssymb,
    floats,
    superscriptaddress,
    nofootinbib,
    showpacs
]{revtex4-2}

\usepackage{multirow}
\usepackage{graphicx}
\usepackage{sidecap}
\usepackage[normalem]{ulem}
\usepackage{diagbox}
\usepackage{xcolor}

\usepackage{subcaption}
\DeclareCaptionJustification{justified}{\justifying}
\usepackage{hyperref} 

\begin{document}
\author{Daniel~Lanchares}
\email{daniellanchares@gmail.com}
\affiliation{Departamento de Fisica, Universidad de Oviedo, C. Federico Garcia Lorca 18, 33007 Oviedo, Spain}
\affiliation{Instituto Universitario de Ciencias y Tecnologías Espaciales de Asturias (ICTEA), C. Independencia 13, 33004 Oviedo, Spain}
  
\author{Osvaldo Gramaxo Freitas}
\email{osgrade@alumni.uv.es}
\affiliation{Departamento de
  Astronom\'{\i}a y Astrof\'{\i}sica, Universitat de Val\`encia,
  Avinguda Vicent Andrés Estellés 19, 46100, Burjassot (Val\`encia), Spain}
\affiliation{Centro de F\'{\i}sica das Universidades do Minho e do Porto (CF-UM-UP), Universidade do Minho, 4710-057 Braga, Portugal}

\author{Lysiane Mornas}
\email{mornas@uniovi.es}
\affiliation{Departamento de Fisica, Universidad de Oviedo, C. Federico Garcia Lorca 18, 33007 Oviedo, Spain}
\affiliation{Instituto Universitario de Ciencias y Tecnologías Espaciales de Asturias (ICTEA), C. Independencia 13, 33004 Oviedo, Spain}

\author{José A. Font}
\email{j.antonio.font@uv.es}
\affiliation{Departamento de
  Astronom\'{\i}a y Astrof\'{\i}sica, Universitat de Val\`encia,
  Avinguda Vicent Andrés Estellés 19, 46100, Burjassot (Val\`encia), Spain}
\affiliation{Observatori Astron\`omic, Universitat de Val\`encia,  Catedr\'atico 
  Jos\'e Beltr\'an 2, 46980, Paterna (Val\`encia), Spain}

\author{Joaqu\'{\i}n González-Nuevo}
\email{gnuevo@uniovi.es}
\affiliation{Departamento de Fisica, Universidad de Oviedo, C. Federico Garcia Lorca 18, 33007 Oviedo, Spain}
\affiliation{Instituto Universitario de Ciencias y Tecnologías Espaciales de Asturias (ICTEA), C. Independencia 13, 33004 Oviedo, Spain}

  \author{Luigi Toffolatti}
\email{ltoffolatti@uniovi.es}
\affiliation{Departamento de Fisica, Universidad de Oviedo, C. Federico Garcia Lorca 18, 33007 Oviedo, Spain}
\affiliation{Instituto Universitario de Ciencias y Tecnologías Espaciales de Asturias (ICTEA), C. Independencia 13, 33004 Oviedo, Spain}

  \author{Pietro Vischia}
\email{vischia@uniovi.es}
\affiliation{Departamento de Fisica, Universidad de Oviedo, C. Federico Garcia Lorca 18, 33007 Oviedo, Spain}
\affiliation{Instituto Universitario de Ciencias y Tecnologías Espaciales de Asturias (ICTEA), C. Independencia 13, 33004 Oviedo, Spain}

\title{Assessment of normalizing flows for parameter estimation on time-frequency representations of gravitational-wave data}

\begin{abstract}
The speed-up of parameter estimation is an active field of research in gravitational-wave data analysis. In this paper we present GP15, a deep-learning method that merges residual networks and normalizing flows into a general-purpose, image-based estimator of binary black hole (BBH) parameters. Building on our early work, we map BBH spectrograms from the Advanced LIGO and Advanced Virgo detectors to color channels in an RGB image amenable to be processed with residual networks. GP15 is trained on simulated data for BBH mergers obtained with the \texttt{IMRPhenomXPHM} waveform approximant and tested for all three-detector events from the GWTC-3 and GWTC-2.1 catalogs reported by the LIGO-Virgo-KAGRA (LVK) collaboration. Overall, our model yields good agreement with the LVK results over most parameters. Our simple model can produce large amounts of posterior samples in the order of a second, complementing existing approaches with normalizing flows based on time or frequency representation of gravitational-wave data. We also discuss current shortcomings of our model and possible improvements for future extensions (e.g.~including noise conditioning from the detectors’ PSD or splitting the parameter space into intrinsic and extrinsic subspaces).
\end{abstract}

\maketitle

\section{Introduction}
\label{sec:intro}

A decade after the momentous observation of gravitational-wave (GW) signal GW150914~\cite{GW150914}, 
GW astronomy remains in the vanguard of experimental physics. The results of the first three observing runs of the LIGO-Virgo-KAGRA (LVK) collaboration are published in three  GW Transient Catalogs (GWTC)~\cite{GWTC1_paper,GWTC-2,GWTC-3}, listing 90 confidently-detected events, all of them corresponding to compact binary coalescences (CBC). In addition, GWTC-2.1~\cite{GWTC-2.1} reports a deeper list of candidate events observed over the first half of the third observing run (O3a). In the last LVK run O4, 254 more significant candidate events have been publicly reported on the GW Candidate Event Database\footnote{\texttt{https://gracedb.ligo.org/superevents/public/O4}}. The release of GWTC-5.0, corresponding to the second part of O4 (O4b), has updated the number of confident detections to 390 observations~\cite{GWTC-5.0}.
While detections of low mass binaries (such as black hole and neutron star (BHNS) or binary neutron star (BNS) mergers~\cite{GW200105_GW200115,GW170817,GW190425}) have occurred in every observation run, most of the recorded events are binary black hole (BBH) mergers. 

The fact that GW from CBC events can be modeled using techniques from analytical relativity and numerical relativity allows to design a detection strategy based on matched filtering. This approach cross-correlates GW triggers to waveform templates (so-called, \textit{approximants}), maximizing the likelihood of the presence of a GW signal in the data collected by the detectors~\cite{LIGOScientific:2019hgc}. The estimation of the source parameters is usually done through Bayesian statistical inference (see~\cite{Parameter_estimation_bible} and references therein). As for the detection problem, this method makes use of a large number of \textit{approximants} that ideally should cover the parameter space of the source as much as possible. Stochastic sampling techniques are applied to evaluate the likelihood functions and posterior distributions of the parameters. Parameter estimation pipelines rely on nested sampling~\cite{nested_sampling, dynesty_paper}, an accurate yet computationally expensive algorithm built on top of the already costly Markov-Chain Monte Carlo (MCMC) algorithm~\cite{MCMC_Metropolis, MCMC_Hastings}, which operates on a timescale of hours (for BBH merger signals) up to days (for longer BNS merger signals) and can require up to millions of waveform evaluations~\cite{GWTC1_bilby,Parameter_estimation_bible}. Both MCMC and nested sampling are routinely used by the LVK collaboration to infer the parameters of CBC signals ever since the very first detection of signal GW150914~\cite{GW150914}.

Despite the many strengths of current statistical approaches, the development of more computationally efficient procedures to assist parameter inference grows increasingly important as the volume of data increases~\cite{Parameter_estimation_bible}. Indeed, the amount of detections is expected to significantly increase during the upcoming O5 run of the LVK collaboration and, especially, when third-generation detectors become operational. In one year of operation, a network consisting of one Cosmic Explorer~\cite{CE} and the Einstein Telescope~\cite{ET} is expected to detect ${\cal O}(10^5)$ BNS mergers~\cite{2021arXiv210909882E,2024PhRvD.110d3013W}. Given these prospects, GW astronomy is expected to benefit from big data handling approaches. Quite naturally, the application of machine learning (ML) techniques as an alternative to Bayesian parameter estimation has been gaining ground recently, not only within GW astronomy~\cite{Parameter_estimation_bible,Cuoco:2020,Benedetto:2023,Stergioulas:2024}, but also in many other fields~\cite{biomediacal_application_ML_estiamtion, DL_bayes_nature, ML_applic_biomass}. Current efforts within the GW context involve the use of Deep Learning (DL) approaches, particularly variational autoencoders, convolutional neural networks, and autoregressive normalizing flows (NFs)~\cite{GW_DL_2018,2018PhRvL.120n1103G,2020PhRvL.124d1102C,2020PhRvD.102j4057G,Base_paper_Valencia,BNS_DL_2021_136161,2022MLS&T...3a5007S,2023PhRvL.130q1403D,tunning_GW_NPE,GW_estimation_CNNs_2024,pisa_close_encounters,2024MLS&T...5a5036G,ML_aided_multi_msgr_search,2025Natur.639...49D}.

In this work we report on a study of parameter inference of BBH mergers using NFs ~\cite{normflows_main_source, MAF_paper, neural_spline_flows_paper}. In contrast to previous investigations where NFs were applied on individual time or frequency GW data representations~\cite{2020PhRvD.102j4057G,dingo_real_time_NPE,2023PhRvL.130q1403D,2025Natur.639...49D}, our work explores the performance of NFs using a time-frequency representation of the data. To do so, we closely follow the procedure discussed in Ref.~\cite{Base_paper_Valencia}, where spectrograms from BBH mergers from the Advanced LIGO and Advanced Virgo detectors (L1, H1, and V1~\cite{Advance_LIGO_paper, Advance_Virgo_paper}) were mapped to color channels in an RGB image, suitable to be processed using a Residual Network (ResNet). Our study is also motivated by the work of~\cite{2020PhRvD.102j4057G} where a convincing demonstration of the relevance of NFs for GW parameter estimation was first presented. The initial study of~\cite{2020PhRvD.102j4057G} led to \textsc{Dingo}~\cite{dingo_real_time_NPE}, a highly efficient and accurate algorithm for neural posterior distribution estimation, able to reduce the inference of real LVK events  from ${\cal O}({\rm days})$ to 20 s per event. This algorithm has recently been extended to conduct rapid inferences of both BBH mergers and BNS mergers~\cite{2023PhRvL.130q1403D,2025Natur.639...49D}.
The model we report in this work, dubbed GP15, is trained on spectrograms made from simulated data for BBH mergers obtained with the \texttt{IMRPhenomXPHM} waveform approximant~\cite{IMRPhenomXPHM} and tested on all three-detector events from GWTC-3 and GWTC-2.1. As we show below, our model yields an overall good agreement with the LVK results over the majority of BBH parameters. Moreover, our new model significantly surpasses our original approach~\cite{Base_paper_Valencia} where the inference was done through a combination of ResNets and Monte Carlo dropout. This study proves that NFs can also be applied for efficient and accurate posterior estimations when using time-frequency representations of GW data.

This paper is organized as follows. In Section~\ref{sec:methodology} we outline the methodology of our approach and the main building blocks: image representations of CBC signals, NFs, dataset generation pipeline, model architecture, and training procedure. Section~\ref{sec:results} presents the main results and provides a comparison between samples of our trained model and GWTC data. Finally, our main conclusions are summarized in Section~\ref{sec:conclusions} along with a brief discussion. 

\section{Methodology}
\label{sec:methodology}

\subsection{Stacked spectrogram representation}
\label{subsec:image-DL}

\begin{figure}[t]
      \centering
      \includegraphics[width=0.48\textwidth]{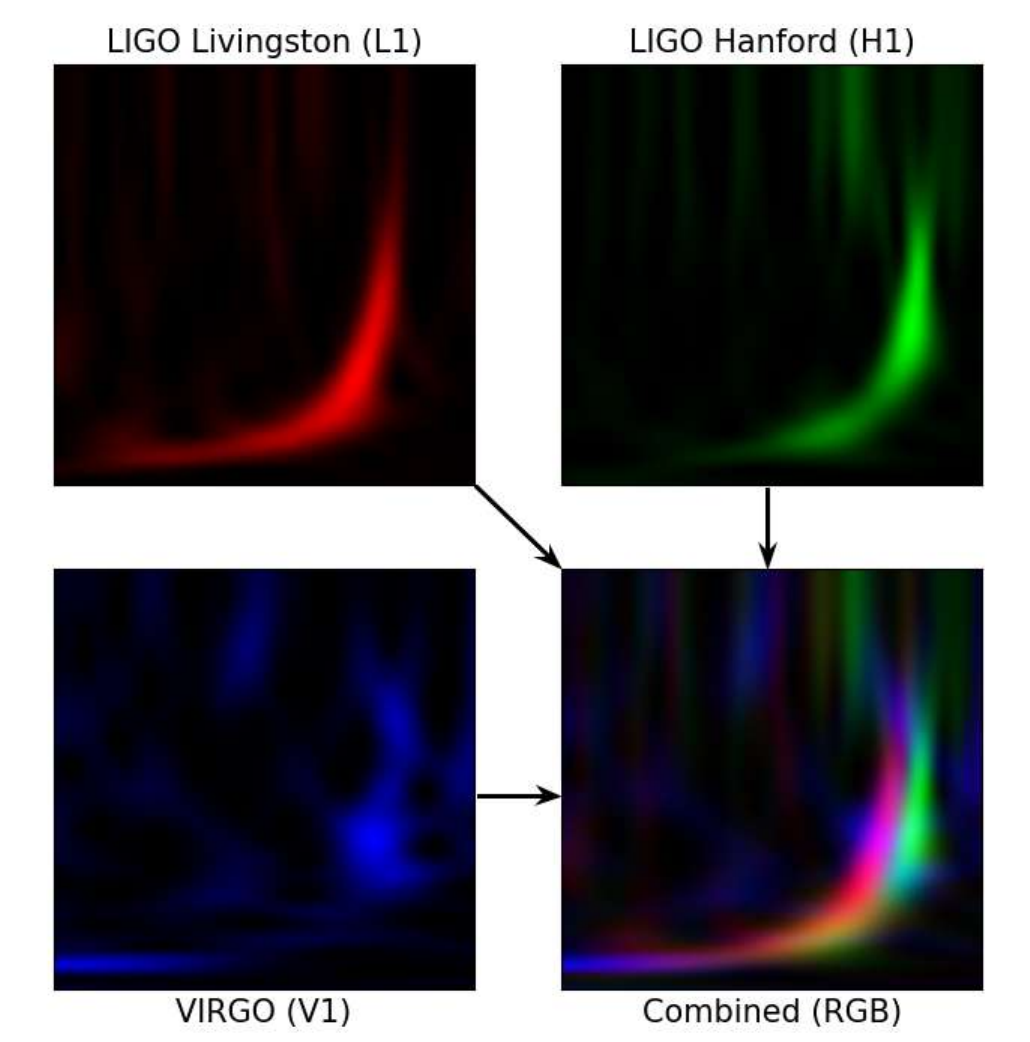}
      \captionof{figure}{Example of time-frequency representation of a loud CBC GW injection, combining data from three detectors (L1, H1, and V1) to form an RGB image.}
    \label{fig:diagram_RGB}
\end{figure}

GW are detected through the measurement of infinitesimal variations in the distance between the test masses located at the extremities of the detector's arms. This data is subsequently encoded into one time series, which is utilized by estimation algorithms when an event (or trigger) is presumed to have been captured. The majority of these algorithms, whether Bayesian in nature or relying on DL methodologies, operate directly on this representation or on its frequency domain counterpart~\cite{bilby_paper}. However, a two-dimensional representation of the GW signal, known as a time-frequency plot or \textit{spectrogram}, can also be constructed and successfully employed for detection and parameter estimation tasks, such as in the coherent WaveBurst (cWB) code~\cite{cWB, cwb_2}. With such a representation, in the context of CBC systems, the rapid rise in frequency of the waveform during the final cycles prior to merger (the so-called \textit{chirp}) carries a large amount of energy and is often distinctly visible in spectrograms. 
In Fig.~\ref{fig:diagram_RGB} we present an example of this representation for a loud GW injection in a three-detector network consisting of LIGO-Livingston, LIGO-Hanford~\cite{Advance_LIGO_paper}, and Virgo~\cite{Advance_Virgo_paper}. Such a configuration is naturally aligned with an RGB image representation, in which each detector is assigned to a distinct color channel. This proves useful for a ML-oriented exploration since image-based learning has always been at the forefront of industrial and medical research, and used across a plethora of fields~\cite{Review_ANNs_industry, Review_CNNs, image-based-radiotherapy-review, transf_learning_malicious_software, transf_learning_chinese_frescoes}. As such we can take advantage of existing performant architectures, such as ResNets~\cite{resnets_paper}, in order to extract features from stacked spectrograms of GW signals (see also~\cite{Base_paper_Valencia} for further details). 

\subsection{Normalizing flows for posterior estimation}
\label{subsec:flows}

Normalizing flows are a class of generative models in ML, typically used for modeling probability distributions~\cite{normflows_main_source}. A NF is composed of a collection of parametrized, invertible, differentiable transformations that convert the samples from an easy-to-sample base distribution into an arbitrarily complex distribution, following a bijective mapping (the \textit{flow}). Since the composition of invertible functions is itself invertible, back-propagation algorithms can be employed to train the parameters of NFs. This allows the optimization of a loss function that quantifies the difference between a target distribution and the model's output: by minimizing this loss function, the model can be fine-tuned to accurately approximate complex probability distributions. Crucially, the flow can be conditioned on additional contextual information, enabling the model to generate samples that are correlated to specific conditioning inputs. Our goal is to obtain posterior parameter distributions of GW signals by conditioning a NF on extracted features from stacked spectrograms of GW detector data.

\subsection{Dataset generation}
\label{subsec:generation_pipe}

\renewcommand{\arraystretch}{1.1}
\begin{table}
   \caption{Default priors for generation of the dataset.
Priors for $\mathcal{M}$ and $q$ are uniform in the mass components, while for $\theta_i$ and $\theta_{JN}$ they are uniform in the sine. The prior for $d_L$ is uniform in comoving volume. It has two upper limits for the two datasets used. Samples are rejected until a signal within the SNR range is created.}
    \centering
    \begin{tabular}{|c|c|}
    \hline
        CBC parameter & Prior \\
    \hline
        $\mathcal{M}$& $\mathcal{U}_{m_1, m_2}\{15, 120\} \mathrm{M_{\odot}}$\\
        $q$& $\mathcal{U}_{m_1, m_2}\{\frac{1}{10}, 1\}$\\
        $a_i$& $\mathcal{U}\{0,0.99\}$\\
        $\theta_i$& $\mathcal{U}_{\sin}\{0, \pi\}$\\
        $\phi_{jl}$& $\mathcal{U}\{0, 2\pi\}$\\
        $\phi_{12}$& $\mathcal{U}\{0, 2\pi\}$\\
 $d_L$&$\mathcal{U}_{cov}\{0.1, 5(10)\} \mathrm{Gpc}$\\
       $\theta_{JN}$& $\mathcal{U}_{\sin}\{ 0, \pi\}$\\
 $\alpha$&$\mathcal{U}\{0, 2\pi\}$\\
 $\delta$&$\mathcal{U}_{\cos}\{-\pi/2, \pi/2\}$\\
        $\phi$& $\mathcal{U}\{0, 2\pi\}$\\
        $\Psi$& $\mathcal{U}\{0, \pi\}$\\
 $t_c$&$\mathcal{U}\{-0.1, 0.1\}$\\
        SNR range&$\left[8, 60\right]$\\
  \hline
    \end{tabular}
    \label{tab:priors}
\end{table}
\renewcommand{\arraystretch}{1}

In order to generate the amount of data we need to train such a network, we turn to waveform approximants, in particular \texttt{IMRPhenomXPHM}~\cite{IMRPhenomXPHM}. In this work, we aim to accurately estimate all the parameters of a CBC event, namely  the detector-frame chirp mass ($\mathcal{M}$), mass ratio ($q$), dimensionless spin amplitudes ($a_1$, $a_2$), their zenith angles ($\theta_1$, $\theta_2$) and their azimuthal differences ($\phi_{JL}$, $\phi_{12}$), luminosity distance ($d_L$), inclination angle of the total angular momentum of the system $\vec{J}$ and the line of sight ($\theta_{JN}$), sky position ($\alpha$, $\delta$), orbital phase ($\phi$), polarization angle ($\psi$), and coalescence time ($t_c$)~\cite{Parameter_estimation_bible}.  The priors used to sample the relevant parameters are reported in Table~\ref{tab:priors}.

After generating the GW waveforms, we make use of utilities from the \textsc{Bilby} library~\cite{bilby_paper} to create six-second Gaussian noise samples based on a pool of PSDs generated from four 500 second long segments of real noise from each of the three detectors and we inject our waveforms into the noise samples. The noise segments start at different O3 GPS times (1240194342, 1249784742 , 1267928742, and 1268431194.1) in order to better mimic the varying noise conditions over an observation campaign.
To ensure the generated data is loud enough to be interpreted, we compute the SNR of our synthetic signals and reject those that are not between 8 and 60, in line with the range of existing LVK detections. If a signal is rejected, we resample the prior and repeat. The resulting timeseries are used to produce stacked spectrograms using signal and image analysis utilities from the libraries \textsc{gwpy}~\cite{gwpy} and \textsc{scikit-image}~\cite{scikit-image} respectively. In particular, we employ a constant-Q transform~\cite{Q-transform} to obtain spectrograms from the whitened injection timeseries for each detector. These are then resampled into three-channel $170\times90$ pixel images in eight-bit color space. Our dataset consists of three million of these images, with 10\% reserved for validation, along with the respective physical parameters for each. An additional three million images were used on the last training stage, for a combined six million data-points to learn from.

\subsection{Model architecture and training}
\label{subsec:modelling}

\begin{figure}[t]
      \centering
      \includegraphics[width=0.48\textwidth]{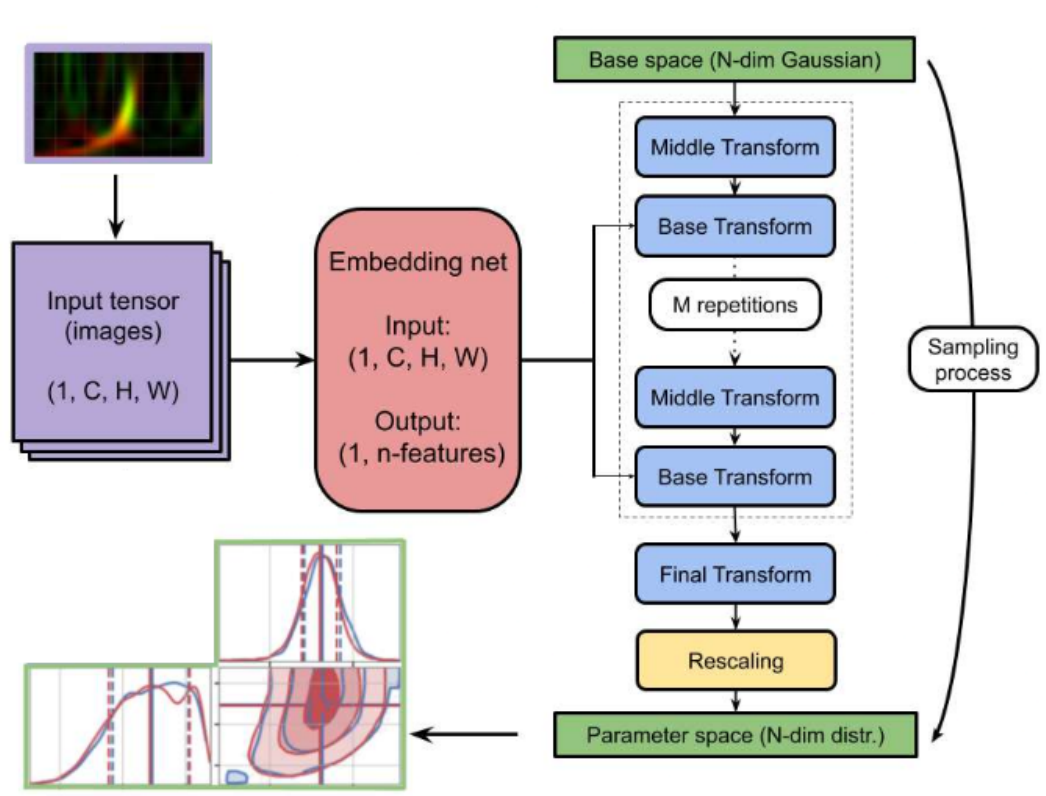}
      \caption{Model diagram depicting the flow of information in the sampling process. 
      The images need to be of shape (channels, height, width) to accommodate the ResNet architecture, which transforms them into a feature vector of length \textit{n-features}. The core of the flow is repeated 25 times, giving a total of 56 million trainable parameters.
      }
      \label{fig:base_model}
\end{figure}

\begin{figure}[b]
      \centering
      \includegraphics[width=0.48\textwidth]{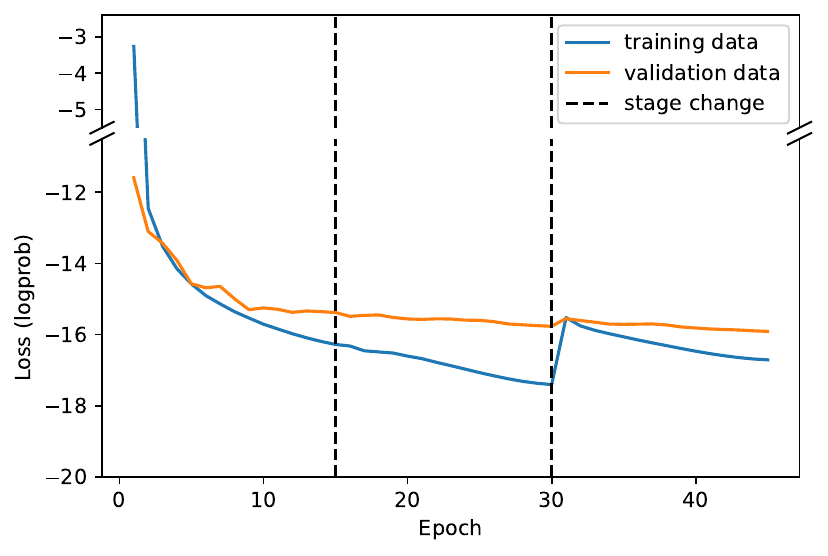}
      \caption{Loss function output averaged over each epoch for both the training (90\%) and validation (10\%) partitions of our dataset. The dashed lines separates the three stages of training.}
      \label{fig:loss_plot}
\end{figure}
\begin{figure}[b]
          \centering
          \includegraphics[width=0.48\textwidth]{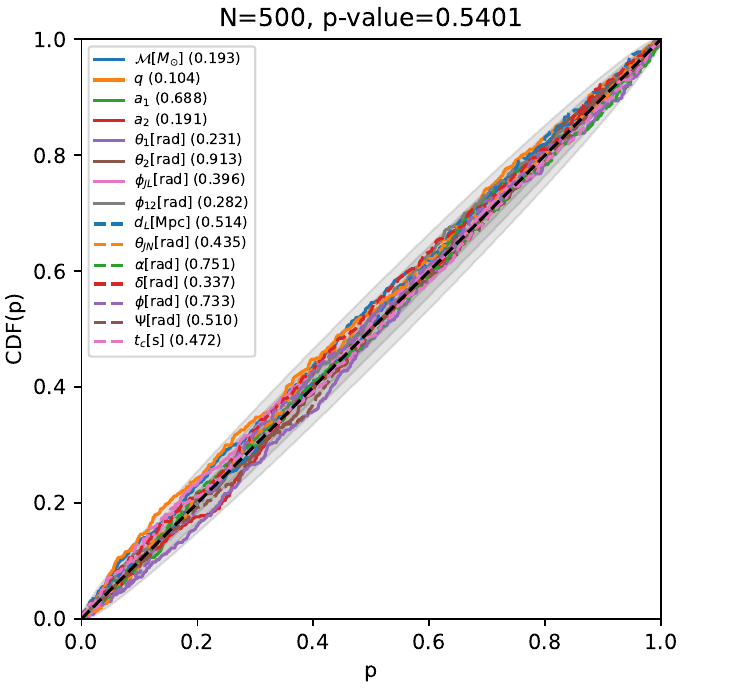}
          \caption{P–P plot for 500 injections with a combined p-value of 0.54. The p-values of the individual parameters are shown in the legend.}
            \label{fig:pp-plot}
\end{figure}
We propose a general-purpose NF model that outputs posteriors for 15 GW parameters (hence its naming, \textsc{GP15}), conditioned by GW feature data. As a base transform, we use a rational-quadratic coupling transform~\cite{dingo_GW150914, neural_spline_flows_paper} followed by a masked affine autoregressive transform~\cite{MAF_paper}. For the middle and final transforms, we combine a random permutation with a LULinear transform, that is, a linear transformation whose matrix of weights is parametrized by the use of a lower-upper (LU) decomposition. We use a ResNet-18~\cite{resnets_paper} as an embedder of GW feature information from stacked spectrogram images.  To regularize training behavior, the target parameters are scaled by the bounds of their priors to map them to the $[0,1]$ interval. An exception to this is the luminosity distance, whose scaling factor of 5000 Mpc maps to the interval for the original dataset, but not for the long distance refining dataset. A diagram of the flow of data through this model during inference can be found in Fig.~\ref{fig:base_model}. We realize this model with the PyTorch framework~\cite{pytorch_paper}, using \textsc{nflows}~\cite{nflows_paper} for the implementation of normalizing flows. 

\begin{figure*}[tbh!]
          \centering
          \includegraphics[width=0.96\textwidth]{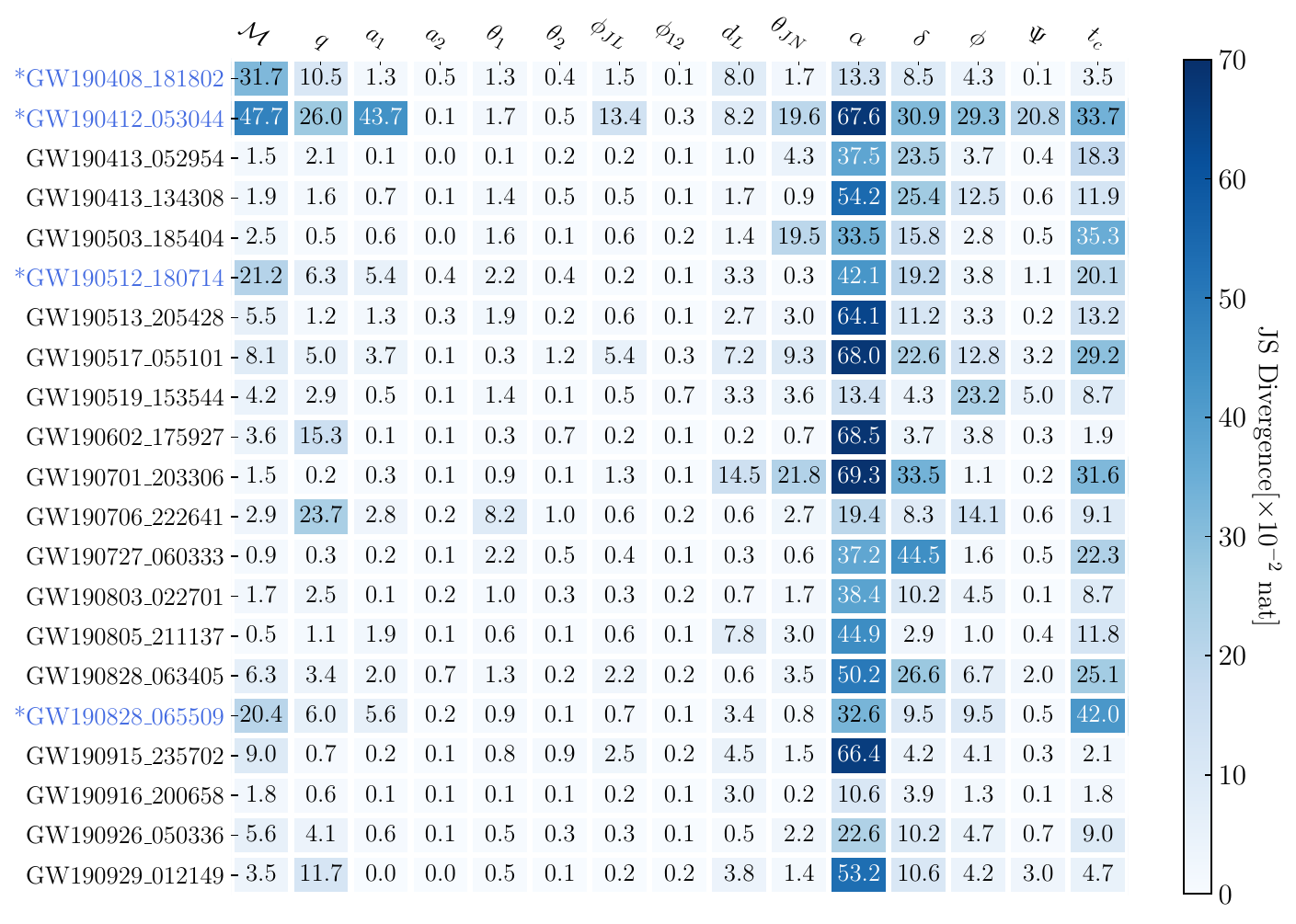}
          \caption{Jensen-Shannon divergences for all three-detector events of GWTC-2.1~\cite{GWTC-2.1} with posterior distributions matching the conditions $\mathcal{M}\gtrsim 15\mathrm{M_{\odot}}$ and $d_L \lesssim 10\mathrm{Gpc}$, dictated by our model's prior. Events marked ($*$) in blue have estimated probability densities below the $15\mathrm{M_{\odot}}$ limit. 10000 samples were generated for each event.}
            \label{fig:GP12_GWTC2}
\end{figure*}

Training is carried out in a three-stage process, in which we employ the \textsc{Adam} optimizer~\cite{Adam_2017} and a cosine annealing strategy~\cite{Cosine_annealing_and_warm_restarts} in order to minimize the loss function,
\begin{align*}
L (d,\tau)&= -\frac{1}{N} \sum_{i=1}^{N}  \log \left| \det \left( \frac{\partial f_\beta(d^{(i)},\tau^{(i)})}{\partial \tau^{(i)}} \right) \right|   \\
&  ~~~~ -\frac{1}{N}\sum_{i=1}^{N}\log p_Z\left(f_\beta^{-1}(d^{(i)},\tau^{(i)})\right)\\
&= - \frac{1}{N} \sum_{i=1}^{N} \log q_\beta(\tau^{(i)}|d^{(i)} ) ,
\end{align*}
where $N$ is the batch size, $f$ is the flow, $\beta$ are the model parameters, $d^{(i)}$ is the ResNet's embedding output, $\tau^{(i)}$ is the set of target physical parameters after scaling, and $p_Z$ is the base distribution. Minimizing $q_\beta(\tau^{(i)}|d^{(i)} )$ over the joint $(\tau, d)$ parameter space is equivalent to minimizing its entropy relative to an ideal Bayesian posterior $p(\tau^{(i)}|d^{(i)} )$. In the first stage of training, we train all the parameters of the model (embedder and flow) up to an intermediate learning rate of $2\times 10^{-5}$ for 15 epochs. The second stage refines the model on the same dataset, annealing the learning rate to 0 over 15 epochs. Initial estimations showed poor performances on medium distance mergers, especially in GWTC-3. To correct these deficiencies, an additional 15-epoch refinement stage was performed using a new dataset in which the luminosity distance maximum is twice the previous one, annealing the learning rate from $1.5\times 10^{-5}$ to 0. The evolution of the loss function during training can be seen in Fig.~\ref{fig:loss_plot}. The average training loss decreases monotonically within the stages, an indication of meaningful and constant learning. The validation loss evolves similarly and settles marginally above the former, indicating that our NF is capable of generalization beyond its training data. A noticeable bump can be appreciated between stages 2 and 3. This bump is to be expected, given that we are subjecting the model to data that were not only new to it but that also came from a previously unexplored section of the parameter space. The training lasted 36 hours per stage on an NVIDIA A100 from the Artemisa Computer Cluster at the University of Valencia\footnote{https://artemisa.ific.uv.es/web}, using batch sizes of 2048 elements. 

\section{Results}
\label{sec:results}

  \begin{figure*}[tbh!]
          \centering
          \includegraphics[width=0.96\textwidth]{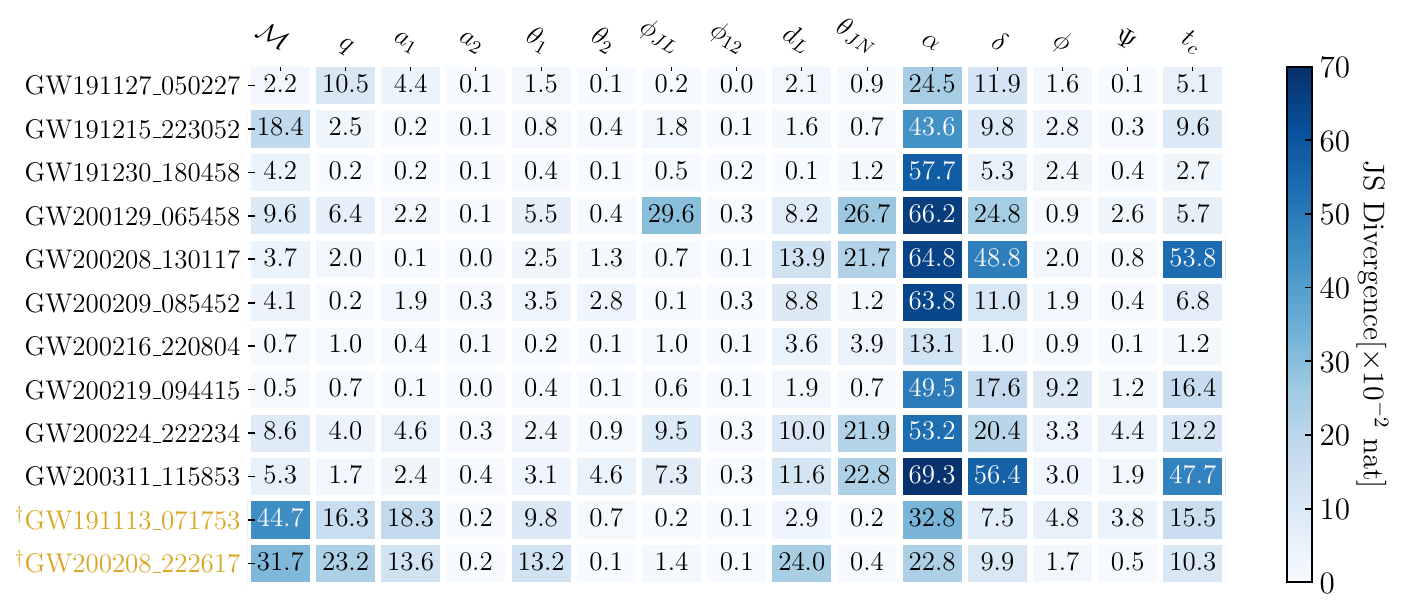}
          \caption{Jensen-Shannon divergences for all three-detector events of GWTC-3~\cite{GWTC-3} with posterior distributions matching the conditions $\mathcal{M}\gtrsim 15\mathrm{M_{\odot}}$ and $d_L \lesssim 10\mathrm{Gpc}$, dictated by our model's prior. Events marked ($\dagger$) in yellow are special cases that have been presented for completeness. 10000 samples were generated for each event.}
            \label{fig:GP12_GWTC3}
    \end{figure*}

Before comparing the model to observations, we evaluate GP15 on synthetic injections consistent with the training data\footnote{For full estimations of  a representative bunch, see Appx.~\ref{sec:corners}}. We sample posteriors from 500 simulated signals and construct a P-P plot (Fig.~\ref{fig:pp-plot}). Following the methodology of~\cite{dingo_real_time_NPE},  we plot the cumulative distribution function (CDF) of the percentile scores of the true values within the marginalized posteriors. Well-behaved scores should be uniformly distributed (diagonal CDF). Our model obtains a combined p-value of $0.54$, indicating it is behaving as expected.

In order to compare our model's results with the LVK results, we filter the events in GWTC-2.1 and GWTC-3 selecting only those with parameters consistent with our priors, as well as detector configuration (i.e.~only three-detector events with quality data for the two LIGO detectors and Virgo). Additionally, we select the GWTC samples inferred with \texttt{IMRPhenomXPHM}. This yields a total of 33 events. To give a quantitative comparison for all events, we compute the Jensen-Shannon Divergence (JSD)~\cite{js_distance}, a symmetric measure of the similarity of two probability distributions ranging from $0$ to $\ln2\, (\approx 0.693)$. The smaller the JSD the larger the similarity between the compared distributions. 

Figures~\ref{fig:GP12_GWTC2} and~\ref{fig:GP12_GWTC3} show the JSD for the selected GWTC-2.1 and GWTC-3 events, respectively, for all 15 parameters. In general, GP15 performance can be considered satisfactory for most inferred parameters, with a few caveats. When estimating the detector-frame chirp mass, for example, the majority of subpar performances (high JSD values) correspond to events with a median $\mathcal{M}\sim 15\mathrm{M_{\odot}}$, marked in blue in Fig.~\ref{fig:GP12_GWTC2}.
The inference of the mass ratio is highly compatible across events. Spin parameters, together with luminosity distance, phase and polarization angle, show good agreement across the board, with JSD values of the order of $10^{-2}$ to $10^{-3}$. Our model understands the degeneracy in the inclination angle, although it struggles with events where the GWTC catalog is confident in a single mode, such as GW190701 or GW200129.
Sky localization, and particularly right ascension, presents the poorest agreement with the catalogs. This performance is often correlated with subpar estimations of coalescence time, though these are the minority.
Events GW191113\_071753 and GW200208\_222617 were singled out as problematic by other previous analyses. For example, they were excluded from 4-OGC~\cite{Nitz_2023_4-OGC}. GW191113\_071753 also has a median $\mathcal{M}$ well below ($13.24\mathrm{M_{\odot}}$) our lower prior limit for the chirp mass. As for GW200208\_222617, it presents a bimodal chirp mass posterior distribution, which is unusual. This event has recently been signaled as a potential eccentric binary~\cite{Planas_eccentricity_GW200208_22}. However, a comprehensive discussion on the possible explanations of these unusual events goes beyond the scope of the present analysis. Nonetheless, these two merger events offer insights into the behavior of our model in unfamiliar situations.

It is worth highlighting that the generation of a set of 10,000 samples with GP15 lasts only $1.13\pm0.13$ seconds on a NVIDIA A100 GPU, which has to be considered a remarkably short time lapse. As a comparison, the inference with parallel \textsc{Bilby}, exploited to obtain the reference samples used here, can last hours to days to converge, even with 640 CPUs and a phenomenological waveform generator (see Table 1 of Ref.~\cite{parallel_bilby}, to consistently compare our current outcomes with those).

\section{Discussion}
\label{sec:conclusions}

\begin{figure*}[tbh!]
          \centering
          \includegraphics[width=\textwidth]{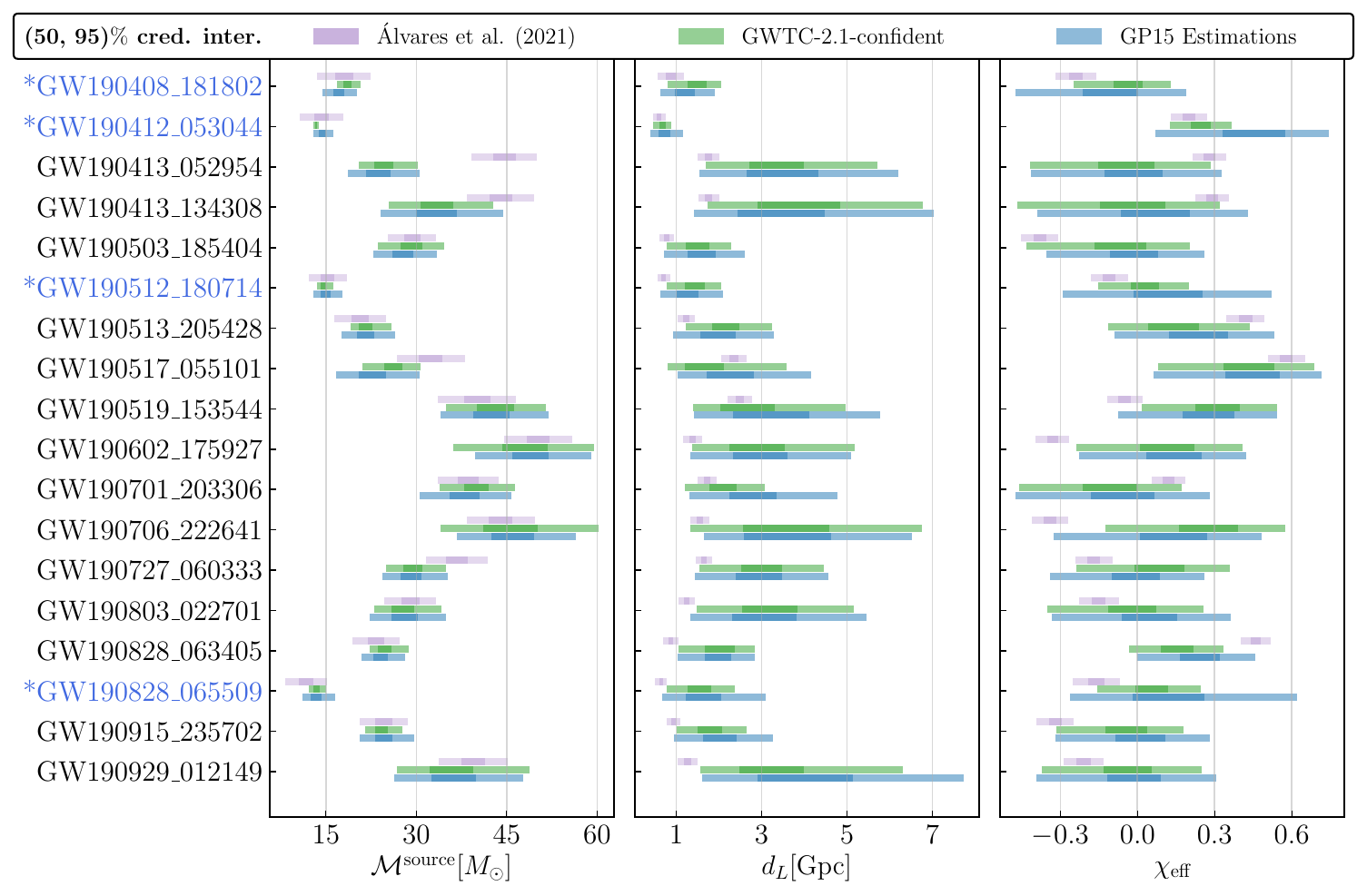}
          \caption{Comparison (50\% and 95\% credible intervals) between the estimations of all compatible events and parameters of \cite{Base_paper_Valencia}. Each three bar group corresponds to a merger. Our earlier work (lilac) on top, GWTC-2.1 reference (green) on the middle and our new model (blue) on the bottom.}
            \label{fig:base_paper_comp}
\end{figure*}

The estimation of the full set of parameters of CBC GW signals, typically achieved through Bayesian statistical approaches in which the posterior inferences are computed with MCMC and nested sampling methods, is a computationally demanding task. Deep Learning approaches such as variational autoencoders, convolutional neural networks, and normalizing flows have been proposed in the last few years to speed up parameter estimation~\cite{Meyer:2022,Parameter_estimation_bible}. In this paper we have joined those efforts by discussing a DL-based method that merges residual networks and normalizing flows into a general-purpose, image-based parameter estimator of CBC systems.  Our model, dubbed GP15, has been trained on spectrograms made from simulated data for BBH mergers obtained with the \texttt{IMRPhenomXPHM} waveform approximant~\cite{IMRPhenomXPHM}. 
This procedure followed the methodology established in our previous work~\cite{Base_paper_Valencia}, where spectrograms of data from each of the three detectors (L1, H1, and V1) were mapped to color channels in an RGB image. In contrast to this previous work, where the full inference was done with ResNets and Monte Carlo (MC) dropout, ResNets were used here to process the stacked spectrogram RGB image into feature vectors, which we then used to condition a normalizing flow. Comparing both approaches, the properties of normalizing flows allow for arbitrarily complex distributions that transcend the necessarily Gaussian nature of MC dropout, and the sampling process is more explicitly Bayesian. Once trained, GP15 was tested on publicly available data from the first three LVK collaboration observing runs~\cite{KAGRA:2023pio} (available on the Gravitational Wave Open Science Center). 

The model has been tested for the three-detector events from GWTC-3 and GWTC-2.1 coherent with our prior (amounting to 33 events) and has shown an overall good agreement with the LVK results (estimated through the Jensen-Shannon Divergence) over the majority of parameters. Poor performances were found mainly in the estimation of the right ascension, declination and coalescence time, which form the localization subspace of parameters, equivalent to the relative time-of-arrival at each of the 3 detectors. We have shown that a relatively simple model such as GP15 can produce large amounts of posterior samples of an unknown target distribution in the order of a second. As already discussed elsewhere (see~\cite{dingo_real_time_NPE,2023PhRvL.130q1403D,2025Natur.639...49D}), the significant inference speed-up of normalizing flows may have a major impact for future observing runs from the LVK and third-generation detectors, since the possibility of parameter estimation in low-latency allows for rapid identification of suitable candidates to conduct multi-messenger follow-up observations.

This work also represents a leap in performance compared to Ref.~\cite{Base_paper_Valencia}. This is shown in Fig.~\ref{fig:base_paper_comp} which displays a comparison of the results presented in Fig.~12 of~\cite{Base_paper_Valencia}, for all events compatible with those used here, along with the estimations of our new model. Once again, we mark events whose posteriors lie near our prior's boundaries. Note that our model does not directly estimate two of the three parameters (source-frame chirp mass and effective spin), so a conversion is performed according to the following expressions: 
\begin{equation} \label{eq:chi_eff}
    \mathcal{M}^{\rm source} = \frac{\mathcal{M}}{1+z} \hspace{1cm} \chi_{\rm eff} = \frac{m_1\chi_1+m_2\chi_2}{m_1+m_2}\,.
\end{equation}
Particular care has to be taken with the source-frame chirp mass (referred to as chirp mass during this comparison) due to the necessity to calculate redshift from luminosity distance and therefore assume a particular set of cosmological parameters. Here we adopt the parameters published in~\cite{Planck_2015_params}, as this set of parameters is the most commonly used in the field of GW cosmology.

Fig.~\ref{fig:base_paper_comp} shows that we achieve a significant improvement on the chirp mass inference with respect to our previous methodology~\cite{Base_paper_Valencia} while being mostly in agreement with GWTC-2.1. Our results reinforce the previous ones where they agreed with GWTC data (e.g.~for GW190503), and correct them where they disagreed (e.g.~for the pair of GW190413 events). This is worth noting, considering that we are using a different signal representation than in the published catalog. It also reinforces the idea that, with enough training data and model expressivity, spectrograms can provide accurate chirp mass posteriors.
The estimation of the luminosity distance represents, by far, the greatest improvement. Our previous methodology had an upper prior limit of $4 \ \mathrm{Gpc}$, which combined with small datasets and SNR-informed rejection of prior samples, greatly disfavored long distance events. Deviations from the medians were also artificially low, which GP15 has corrected. Aside from the upgrade in DL techniques, this improvement was achieved by training the model over 15 more epochs using an all-new dataset with a higher upper prior limit for luminosity distance (from the original $5\  \mathrm{Gpc}$ to $10\ \mathrm{Gpc}$).
Finally, for the effective spin (a mass-weighted linear combination of the spin projections onto the orbital plane; cf.~Eq.~\ref{eq:chi_eff}), GP15 produces far more realistic posteriors, with credible intervals much more in line with GWTC-2.1.
It is interesting to note that for the events that were marked for their mass, effective spin estimations are noticeably less confident than their published counterparts, but they still represent a major improvement over our previous work. 
In summary, the architectural and computational upgrade in our methodology compared with our previous investigation~\cite{Base_paper_Valencia} has significantly improved results. It also provides further evidence in support of machine-learning, image-based inference pipelines as a model to achieve results comparable to those attained with Monte Carlo approaches.

It is also important to discuss how our findings compare with those obtained using \textsc{Dingo}, the standard benchmark for neural posterior estimation in CBCs. In~\cite{dingo_real_time_NPE}, the authors report an average JSD value with respect to \textsc{Bilby} inferences of  $1.5\times10^{-3}$, with a maximum value of   $1.7\times10^{-2}$ . In contrast, our average value is $8.0$ $(3.6)\times10^{-2}$ with (without) the localization subspace. These differences are significant and deserve some exploration. A potential explanation is that our model, contrary to \textsc{Dingo}, does not include a way to explicitly pass the information of the noise spectrum of each conditioning input, which makes it less flexible when dealing with noise conditions not present in the training set, despite a whitening procedure. In \textsc{Dingo}, the addition of proper noise conditioning leads, for the illustrative case of GW150914, to significantly different posteriors ($3\times10^{-3}$ average JSD with respect to the properly conditioned input, with a maximum of $3\times10^{-2}$). As such, the addition of noise context might improve our results. We must also point to the fact that the model used in \textsc{Dingo} is larger (it contains roughly 3 times more trainable parameters than our model), thereby having more expressivity, and it is trained for a far longer period of 450 epochs, compared to our 45. 
In addition, \cite{dingo_real_time_NPE} also employ an on-the-fly sampling scheme for what they consider extrinsic parameters, namely our localization subspace, polarization angle and luminosity distance. Therefore, their models learn from new sky positions every epoch. Extensions of our methodology will prioritize the incorporation of this two-stage parameter sampling scheme and be reported elsewhere. An explicit treatment of PSD data would be expected to improve results, but it is not as well suited for image representations as detector data itself.

The localization subspace also has symmetries we are not directly accounting for. While a convolutional embedding such as ours might be able to capture this in principle, we had trouble integrating these parameters in a more performant way without severely complicating the architecture of our model. Some improvements were obtained by employing a rectangular image window that prioritizes temporal resolution in the spectrogram representation, as opposed to a square window, but that comes at the expense of frequency resolution. We nevertheless provide users with the option to enable such a window in the code released with this manuscript. We note that \textsc{Dingo} uses a separate neural network to calculate an initial guess for the polarization angle and luminosity distance,  iteratively time-shifting the data to a standard position during inference~\cite{dingo_GNPE}. 

On top of that, we have to take into account the fact that spectrograms inherently lack phase information. This may be relevant to some parameters such as sky location and inclination, though it is not particularly relevant given Virgo's current sensitivity.

Finally, one must also consider the possibility that using spectrogram data may yield different results than using time-series data. Simulation-based inference in a one-dimensional time or frequency domain representation amortizes the Bayesian process of finding the parameters that produce the waveforms that best match the 1D data, which is precisely what \textsc{Bilby} posteriors represent. On the other hand, when using the time-frequency representation one finds the parameters that best reproduce the respective spectrograms. In the presence of colored noise, the correspondence between these processes, while certainly correlated, is not fully ensured. 
Since we have found that vital information such as mass, spin or distance can be accurately extracted from spectrogram data alone, we expect hybrid models trained on a 1D + 2D representation of the data to improve performance by bridging the gap between the current approaches. This multi-head structure would facilitate upgrades such as PSD conditioning. Such models may prove useful for estimation in the presence of non-stationary noise~\cite{Sun_2024_glitch_comb_repr}.

This work demonstrates that our goal of performing well-behaved parameter estimation directly on time–frequency representations of GW events has been achieved, with significantly improved robustness compared to our previous work~\cite{Base_paper_Valencia}. Aside from allowing an alternative interpretation of the data, the methodology and the code discussed here open the door to applications such as multi-band analysis of sources with particularly complex or less well-modeled frequency evolutions. Exploring the use of sparse time-frequency representations, such as the ones used in cWB, might allow for a more flexible representation, helping to expand the class of GW sources our method might be applied to. 
Finally, a natural extension of this work is to study how glitches affect inference in time–frequency methods compared to one-dimensional approaches.

The code that was developed for this project, including a detailed example, is publicly available at \href{https://github.com/Daniel-Lanchares/dtempest}{https://github.com/Daniel-Lanchares/dtempest}.

\subsection*{Acknowledgements}

The authors would like to thank the anonymous referee for the careful reading of the original version of the manuscript and for their insightful remarks. 

OGF is supported by the Portuguese Foundation for Science and Technology (FCT) through doctoral scholarship UI/BD/154358/2022. He further acknowledges financial support by CF-UM-UP through Strategic Funding UIDB/04650/2020. JAF is supported by the Spanish Agencia Estatal de Investigación (grant PID2024-159689NB-C21) funded by MICIU/AEI/10.13039/501100011033 and by FEDER / EU, by the Generalitat Valenciana (CIPROM/2022/49), and by the European Horizon Europe staff exchange (SE) programme HORIZON-MSCA-2021-SE-01 (NewFunFiCO-101086251). JGN acknowledges the projects PID2021-125630NB-I00 and CNS2022-135748 funded by MCIN/AEI/10.13039/501100011033/FEDER, and by the EU “NextGenerationEU/PRTR”. LT acknowledges the Spanish Ministerio de Ciencia, Innovación y Universidades for partial financial support under the projects PID2022-140670NA-I00 and PID2021-125630NB-I00. PV is supported by the “Ramón y Cajal” program under Project No. RYC2021-033305-I funded by the MCIN MCIN/AEI/10.13039/501100011033 and by the European Union NextGenerationEU/PRTR, and by the European Innovation Council (EIC) Pathfinder project PHINDER, grant agreement No. 101258353, funded by the European Union. The authors gratefully acknowledge the computer resources at Artemisa and the technical support provided by the Instituto de Fisica Corpuscular, IFIC (CSIC-UV). Artemisa is co-funded by the European Union through the 2014-2020 ERDF Operative Programme of Comunitat Valenciana, project IDIFEDER/2018/048. 

Our work has made use of data, software and web tools obtained from the \href{https://www.gw-openscience.org/ }{Gravitational Wave Open Science Center}. We use \texttt{PyCBC}~\cite{pycbc}, \texttt{bilby}~\cite{bilby_paper}, and \texttt{PESummary}~\cite{pesummary} for the generation of datasets and the acquisition and handling of GW data. The neural networks are implemented in \texttt{PyTorch}~\cite{pytorch_paper}, and the normalizing flow is constructed from transform modules implemented in \texttt{nflows}~\cite{nflows_paper} loosely following the structure of \textsc{Dingo}~\cite{dingo_real_time_NPE}. Plots are created using \texttt{matplotlib}~\cite{Matplotlib}.
The open source nature of the GW-science community made this research possible.

This material is based upon work supported by NSF’s LIGO Laboratory which is a major facility fully funded by the National Science Foundation, as well as the Science and Technology Facilities Council (STFC) of the United Kingdom, the Max-Planck-Society (MPS), and the State of Niedersachsen/Germany for support of the construction of Advanced LIGO and construction and operation of the GEO600 detector. Additional support for Advanced LIGO was provided by the Australian Research Council. Virgo is funded, through the European Gravitational Observatory (EGO), by the French Centre National de Recherche Scientifique (CNRS), the Italian Istituto Nazionale di Fisica Nucleare (INFN) and the Dutch Nikhef, with contributions by institutions from Belgium, Germany, Greece, Hungary, Ireland, Japan, Monaco, Poland, Portugal, Spain. KAGRA is supported by Ministry of Education, Culture, Sports, Science and Technology (MEXT), Japan Society for the Promotion of Science (JSPS) in Japan; National Research Foundation (NRF) and Ministry of Science and ICT (MSIT) in Korea; Academia Sinica (AS) and National Science and Technology Council (NSTC) in Taiwan.

\bibliographystyle{apsrev}
\bibliography{main} 

\appendix
\begin{figure*}[tbh!]
    \includegraphics[width=\textwidth]{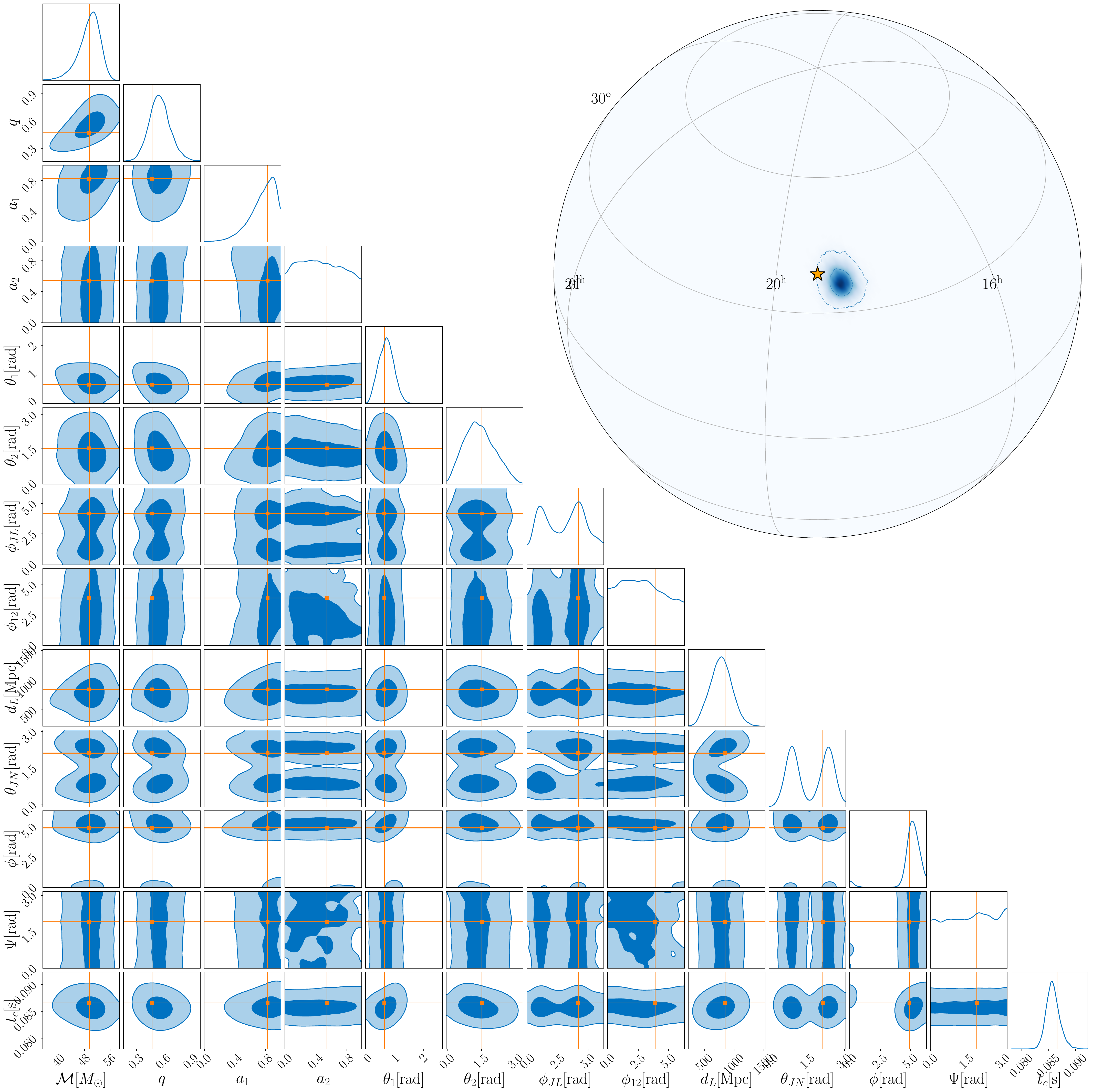}
    \caption{Posterior distributions (blue, 50\% and 95\% credible interval) for a high-scoring synthetic signal. Injected values are denoted in orange.}
    \label{fig:inj-best}
\end{figure*}
\begin{figure*}
    \includegraphics[width=\textwidth]{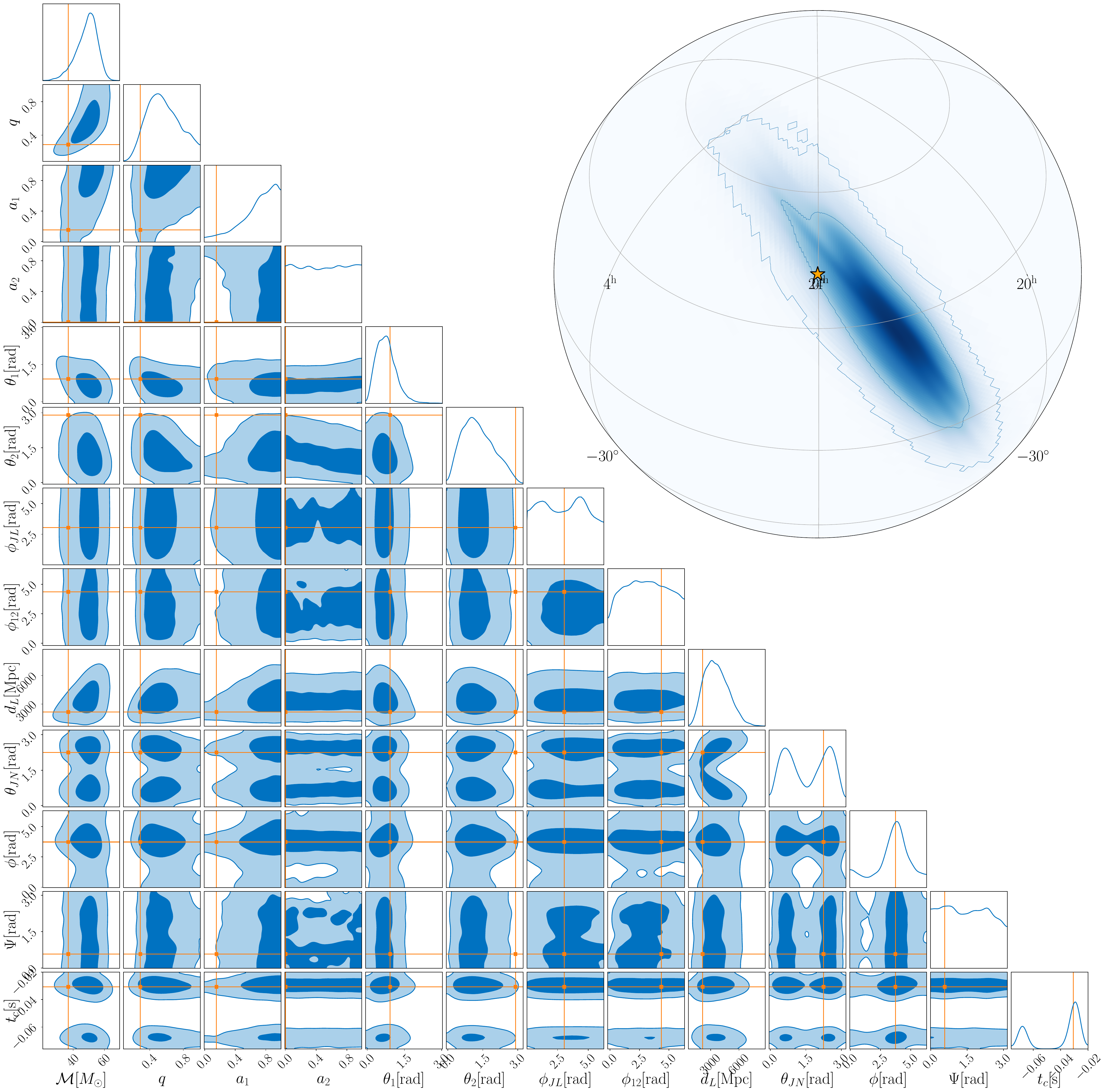}
     \caption{Posterior distributions (blue, 50\% and 95\% credible interval) for a low-scoring synthetic signal. Injected values are denoted in orange.}
     \label{fig:inj-worst}
\end{figure*}
\section{Full estimation of a representative collection of signals}\label{sec:corners} 
We present the marginalized posterior distributions generated by GP15 for four representative signals, corresponding to both synthetic injected data and actual GWTC events. For each category, we include a "best" and a "worst" case scenario to contextualize Figs.~\ref{fig:GP12_GWTC2} and~\ref{fig:GP12_GWTC3}. To quantify this, we ranked simulated data by their mean squared error with respect to injected values averaged over all parameters and the real, unmarked, events by the averaged JSD values, with special attention to deviations in the chirp mass, given its relative importance for this representation of the data.

Fig.~\ref{fig:inj-best} corresponds to the best scoring injection from a pool of 500 generated by sampling the prior described in Tab.~\ref{tab:priors}. Parameters with uni-modal distributions are correctly estimated, and so are bi-modals, up to degeneracy. Sky position is perhaps the least performant (coherent with Figs.~\ref{fig:GP12_GWTC2} and~\ref{fig:GP12_GWTC3}), yet it is still within the 90\% credible interval in spite of a very informative posterior.

Fig.~\ref{fig:inj-worst}, on the other hand, misses the mark on many of the most relevant parameters. Both chirp mass, mass ratio, spin magnitudes and distance are underestimated to various degrees. Inclination, phase and coalescence time are still successfully recovered, and the injected sky position falls within the 50\% credible interval. However, this is due in part to a much more uninformative posterior compared to Fig.~\ref{fig:inj-best}. As Fig.~\ref{fig:pp-plot} illustrates, such poor performances are uncommon.
\begin{figure*}
    \includegraphics[width=\textwidth]{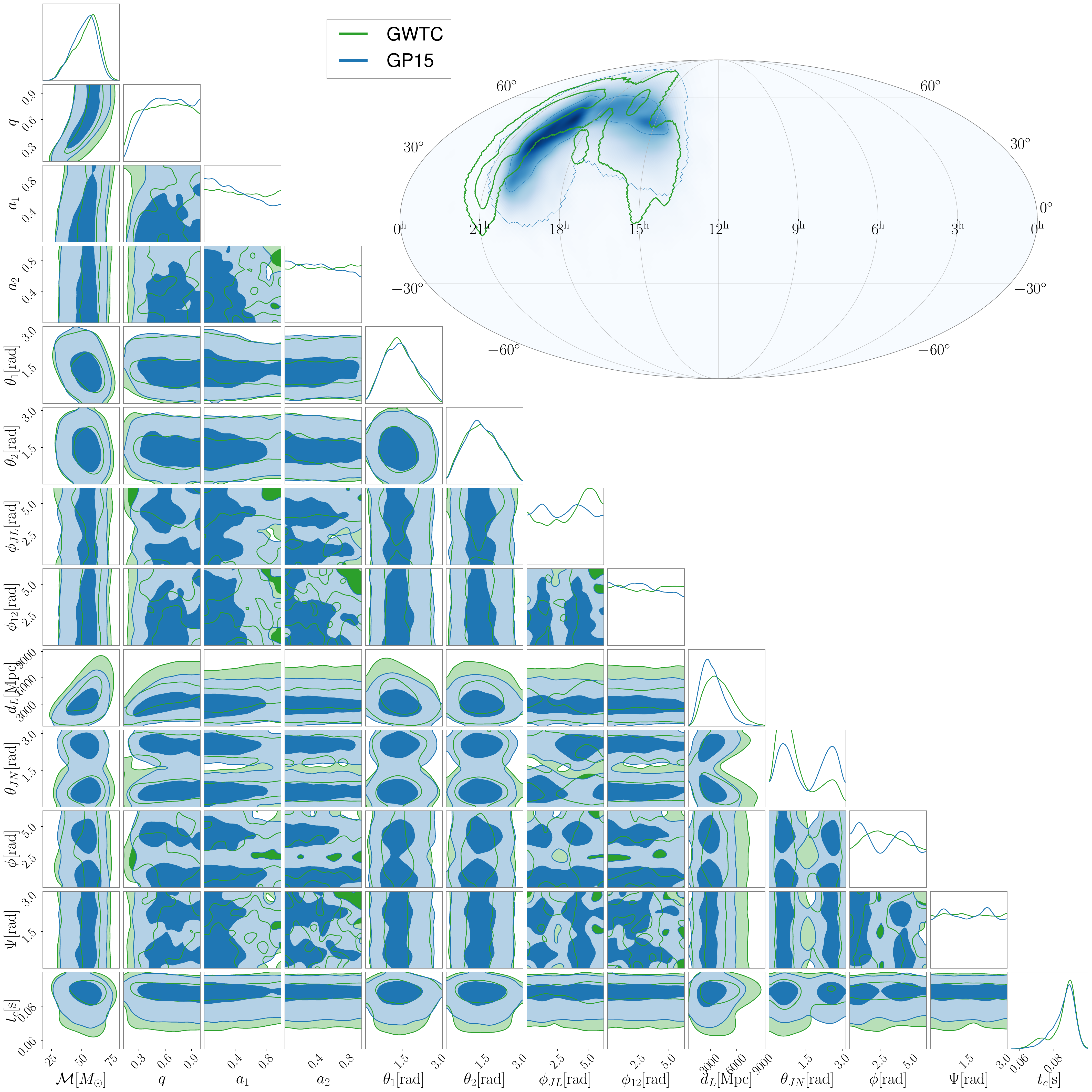}
     \caption{Posterior distributions (50\% and 95\% credible interval) for GW200216\_220804. The posterior distribution estimated by our model appear in blue, while GWTC-3 samples appear in green.}
     \label{fig:LVK-best}
\end{figure*}

For real LVK data, Fig.~\ref{fig:LVK-best} corresponds to GW200216\_220804, a GWTC-3 event. Our ML model estimates posterior distributions that are very similar to those obtained by the nested sampling methods that produced the results published in the different GWTC catalogs~\cite{bilby_paper}. Its biggest differences are its lack of preference for any of the two inclination modes and the overall stellar position, though it is worth noting that it covers a very similar patch of the sky.
\begin{figure*}
    \includegraphics[width=\textwidth]{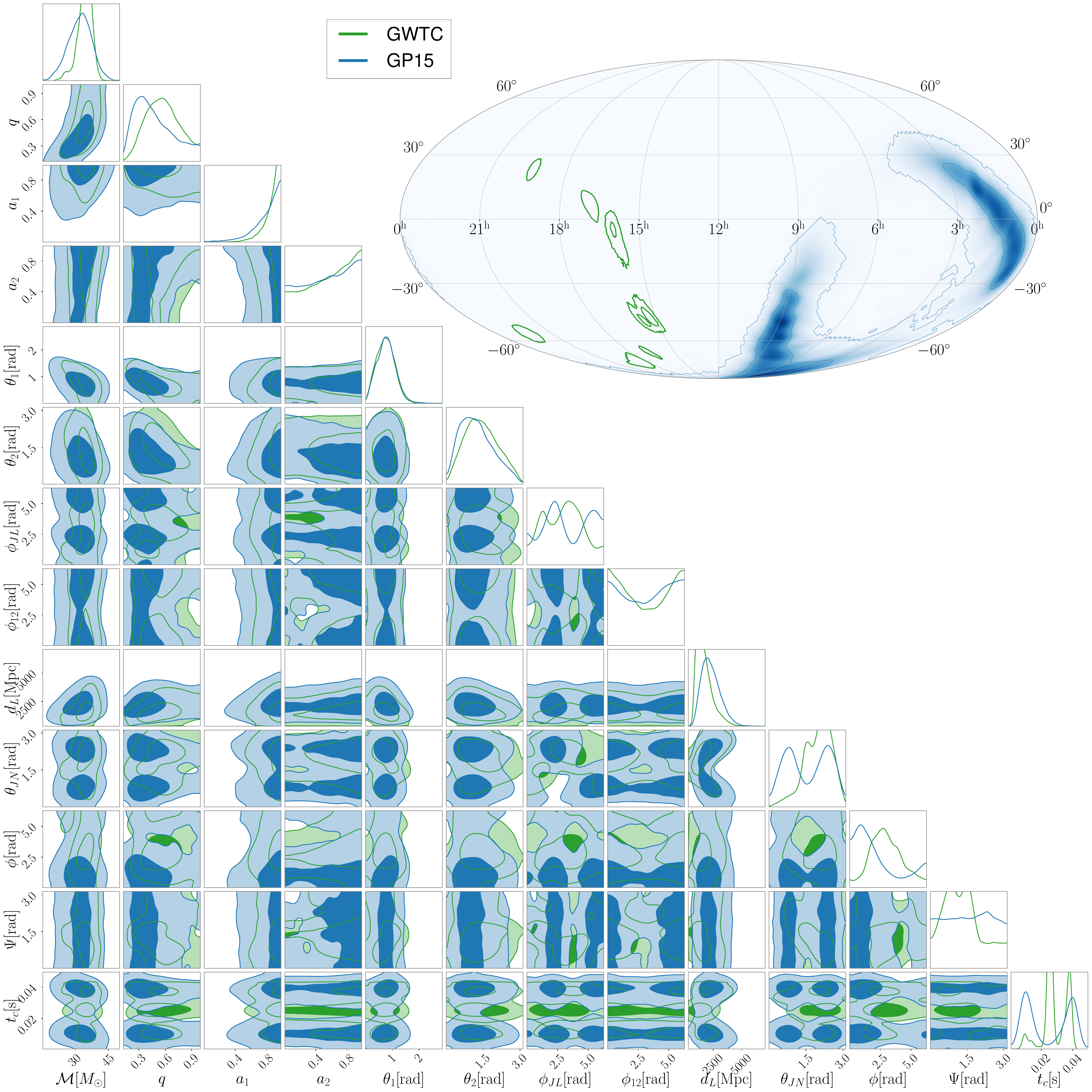}
     \caption{Posterior distributions (50\% and 95\% credible interval) for GW190517\_055101. The posterior distributions estimated by our model appear in blue, while GWTC-2.1 samples appear in green.}
     \label{fig:LVK-worst}
\end{figure*}

Finally, Fig.~\ref{fig:LVK-worst} illustrates a worst case scenario among the studied events. Although not the event with the highest average JSD values, it presented notable deviations in both the localization and the mass subspaces, making it a more relevant case study. The latter are relatively minor, but the former are much more substantial.  We again see the strong correlation that exists between the time of coalescence and the sky position, where misaligned modes in $t_c$ lead to a very different estimation of the system's position. Deviations in phase and polarization angle are also observed. However, the model agrees with the catalog on the spin of the event to a surprising degree, given the estimations of the other parameters.

\end{document}